\pdfoutput=1
\documentclass[11pt]{article}

\usepackage[preprint]{acl}

\usepackage{times}
\usepackage{latexsym}

\usepackage[T1]{fontenc}
\usepackage[utf8]{inputenc}

\usepackage{microtype}

\usepackage{inconsolata}

\usepackage{graphicx}

\usepackage{multirow}
\usepackage{url}
\usepackage{amsmath}
\usepackage{amssymb}
\usepackage{cprotect}
\usepackage{enumitem}

\title{Web Page Classification using LLMs for Crawling Support}

\author{%
	\textbf{Yuichi Sasazawa and Yasuhiro Sogawa} \\
	Hitachi, Ltd. Research and Development Group \\ 
	\texttt{\{yuichi.sasazawa.bj, yasuhiro.sogawa.tp\}@hitachi.com}
}

\begin{document}
\maketitle
\begin{abstract}	
	A web crawler is a system designed to collect web pages, and efficient crawling of new pages requires appropriate algorithms. While website features such as XML sitemaps and the frequency of past page updates provide important clues for accessing new pages, their universal application across diverse conditions is challenging. In this study, we propose a method to efficiently collect new pages by classifying web pages into two types, ``Index Pages'' and ``Content Pages,'' using a large language model (LLM), and leveraging the classification results to select index pages as starting points for accessing new pages. We construct a dataset with automatically annotated web page types and evaluate our approach from two perspectives: the page type classification performance and coverage of new pages. Experimental results demonstrate that the LLM-based method outperformed baseline methods in both evaluation metrics.\footnote{The code used in the experiments is available at \url{https://github.com/ckdjrkffz/web-page-classifier}.}
\end{abstract}

\begin{figure}[t]
	\centering
	\includegraphics[clip, width=7.4cm]{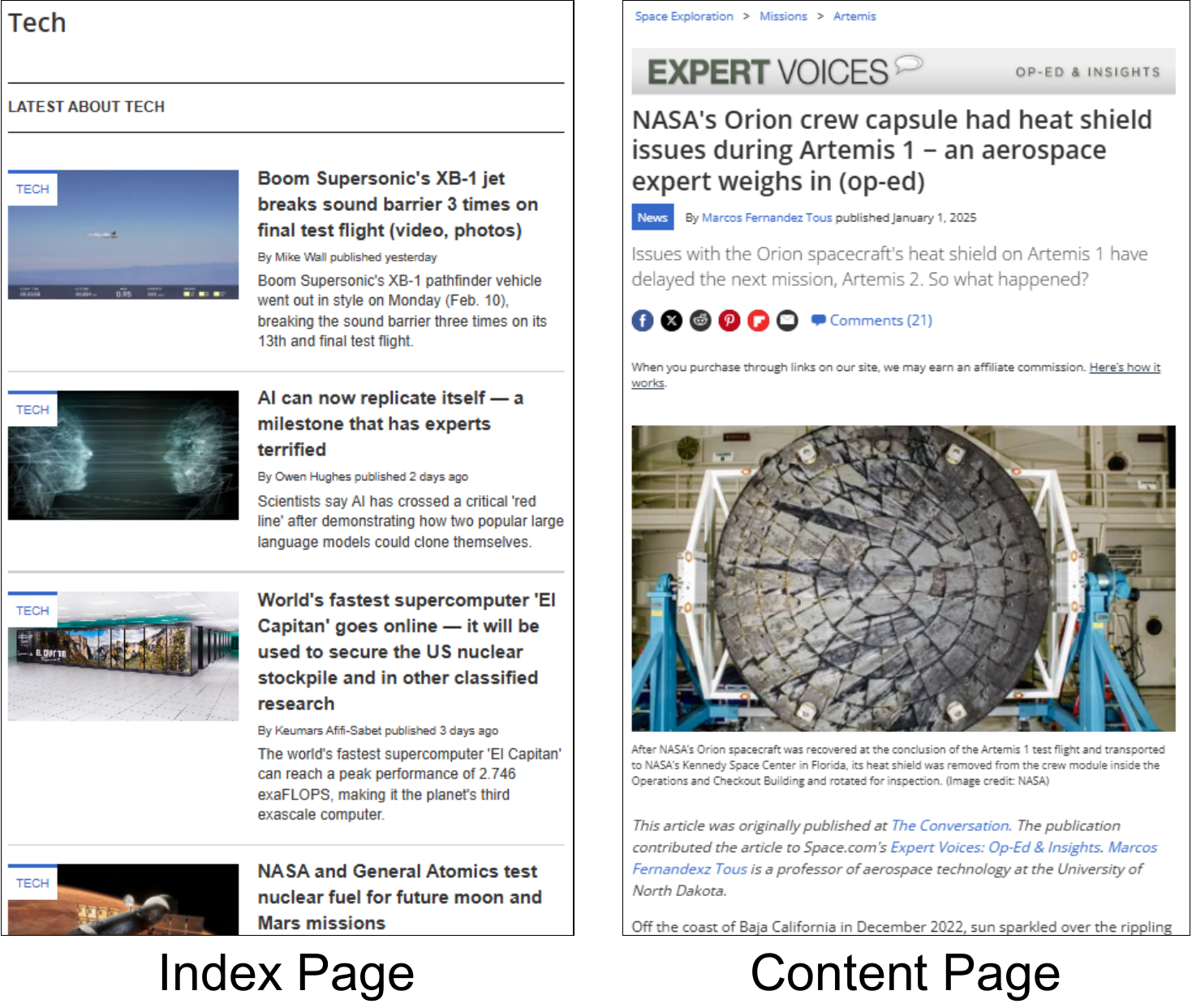}
	\caption{Examples of ``Index Pages''~\cite{space-com-index}, which aim to contain hyperlinks to other pages within the website, and ``Content Pages,''~\cite{space-com-contents} which contain content such as news articles and columns.}
	\label{fig:page_type_example}
\end{figure}

\section{Introduction}
\label{sec:introduction}
A web crawler is a system designed to collect web pages, primarily used to register web pages in search engines~\cite{olston2010web}. As the number of web pages grows daily, efficient crawling of new pages requires the use of appropriate algorithms~\cite{10.1023/A:1019213109274, 10.1002/spe.587, Pant2004}. Commercial crawlers generally rely on features provided by websites, such as XML sitemaps and RSS feeds, to gather updated information~\cite{10.1145/1526709.1526842}. However, there are websites that lack such features. When accessing new pages without relying on site-specific features, it is efficient to inspect only a few pages, such as the home page, but this approach often misses many new pages. Conversely, by revisiting most pages within a website, including outdated ones, it is possible to maintain the freshness of pages within the site and comprehensively collect URLs of new pages, but doing so frequently is impractical due to network and computational resource limitations~\cite{10.1145/371920.371960, 10.1145/958942.958945}. Additionally, the frequency of past page updates can serve as a crucial indicator for determining the appropriate crawling frequency~\cite{10.1145/775152.775246}, but this approach encounters a cold-start problem that cannot handle new pages that have no crawl history~\cite{10.1145/3308558.3313694}.

One way to improve the efficiency of web page collection is to utilize page types. There are many different types of web pages~\cite{10.1145/1363686.1364247, 10.1145/1459352.1459357}, and from the perspective of crawling new pages, it is highly likely that the URLs of new pages can be efficiently collected by identifying web pages intended to provide links to other pages within the website, such as home pages or feed pages, and then concentrating on accessing those pages. However, due to the significant variations in web page structures~\cite{10.5555/645927.672370, 10.1145/1459352.1459357}, it is difficult to automatically identify these pages using heuristic rules.

In this study, we propose a method to enhance crawling efficiency by classifying web page types using large language models (LLMs) and leveraging the classification results. Specifically, as shown in Figure~\ref{fig:page_type_example}, we broadly categorize web pages into two types: \textbf{Index Pages}, which aim to contain links to other pages, and \textbf{Content Pages}, which aim to present content such as news articles and columns.\footnote{While there are other types of pages, such as login pages and error pages, we assume they are relatively few in number. Thus, in this study, pages other than index pages are collectively treated as content pages.} We then perform binary classification using GPT-4o-mini and GPT-4o~\cite{openai2024gpt4ocard}, taking the title and body of web pages as input. LLMs trained on large-scale datasets possess advanced language processing capabilities and are expected to accurately classify even previously unseen web pages. While existing studies have explored topic classification of web pages~\cite{10350067}, filtering harmful pages~\cite{10.1016/j.comnet.2024.111004}, and analyzing HTML~\cite{gur-etal-2023-understanding, huang-etal-2024-autoscraper} using LLMs, to our knowledge, no prior work has investigated methods for classifying page types using LLMs for crawling support.

In our experiments, we construct a dataset by applying an automatic annotation method to label page types. Using the constructed dataset, we evaluate the page type classification performance and demonstrate that LLMs can classify page types effectively compared to the baseline method. Furthermore, we evaluate the coverage of new pages when using the index pages identified by each method as starting points and show that the LLM-based method can improve performance.

\begin{figure*}[t]
	\centering
	\includegraphics[clip, width=15.8cm]{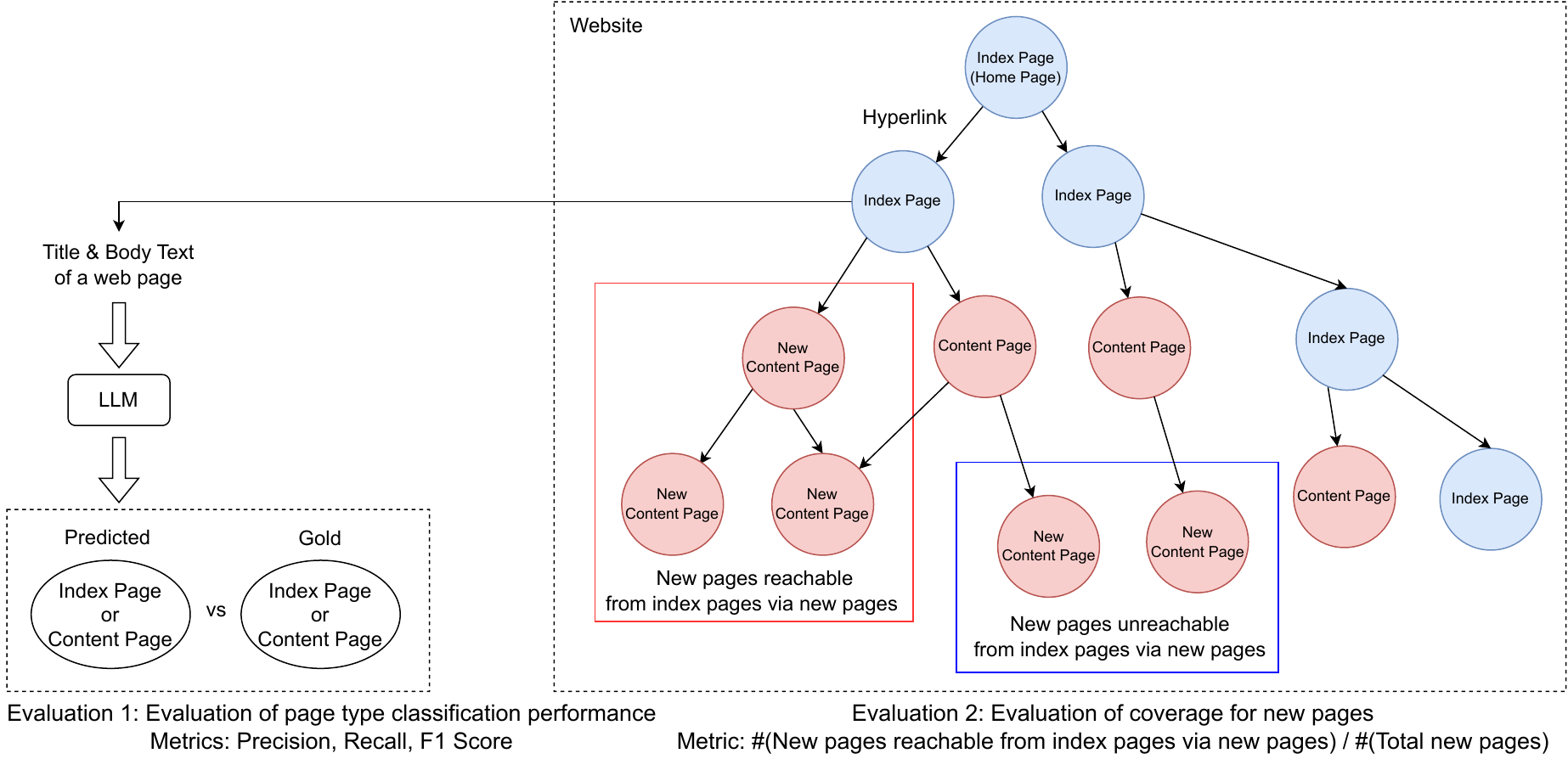}
	\caption{Overview of the proposed method and evaluation approach. Each web page is classified as either an index page or a content page, and new pages are efficiently retrieved starting from index pages. The experiments evaluate two aspects: the page type classification performance and the coverage of new pages.}
	\label{fig:overview}
\end{figure*}

\section{Method}
\label{sec:method}

The overall framework of the proposed method and evaluation approach is illustrated in Figure~\ref{fig:overview}. This study has two main objectives: first, to evaluate the performance of page classification using LLMs, and second, to verify whether classifying page types with LLMs improves the efficiency of retrieving new pages during crawling. In this study, we construct a new dataset annotated with page types. The creation process for this dataset is described in Section~\ref{sec:dataset_creation}. Subsequently, the method for page classification using LLMs is presented in Section~\ref{sec:llm_classification}.

\subsection{Dataset Construction}
\label{sec:dataset_creation}
To evaluate the page type classification performance, a dataset labeled with page types is required. However, such a dataset currently does not exist. Therefore, we construct a new dataset for experimental purposes. Since manually annotating each page would incur high costs, we employ an automated annotation method proposed in this study. The statistics of the constructed dataset are shown in Table~\ref{tab:dataset_statistics}.

\noindent\textbf{Page Type Annotation Method}
Some websites comprehensively list hyperlinks to nearly all content pages on the site across multiple pages, either in the form of XML sitemaps or HTML pages with titles such as ``Latest News'' or ``All Content Archive.'' We refer to these pages as ``Content Listing Pages.'' By classifying the web pages listed on content listing pages as content pages and the pages not listed on content listing pages as index pages, annotation can be performed at a lower cost than manual annotation and potentially closer to human judgment. However, since annotation errors are frequent depending on the website, we perform a simple manual review of the annotation results and retain only those websites deemed to be of sufficient quality.

\noindent\textbf{Detailed Procedure}
We construct the dataset targeting English news websites across multiple domains. First, we select websites that include content listing pages. From these websites, we download 10,000 web pages per site using breadth-first search starting from the home page. Since pages are collected on a site-by-site basis, only internal links are used. Additionally, non-HTML formats such as PDFs are excluded. Next, for each website, we create a script to collect the URLs of content pages linked from the content listing pages, and use these URLs to annotate the type of each page. Finally, we conduct a manual review to retain only websites that meet the quality criteria, and split them into development and test sets on a site-by-site basis.

\noindent\textbf{Construction of the Noisy-Test Set}
For websites with content listing pages, it is expected that the coverage of new pages will be higher than average, as it is easier to access each content page from content listing pages. Therefore, we also download pages from websites that do not have content listing pages using the same procedure and use this as a ``Noisy-Test'' set for evaluation experiments. While this set cannot be annotated with page types and hence is not used for the evaluation of classification performance, it serves as a challenging dataset to evaluate the generality of our method for new page coverage performance.

\noindent\textbf{New Pages Annotation}
The dataset we construct simply captures static snapshots of websites and does not dynamically capture changes to web pages to obtain newly posted pages. Therefore, we treat web pages published within a certain number of days prior to the latest date found among the collected pages of that site as new pages. To assess the performance under multiple crawling frequencies, evaluations are conducted under two scenarios based on the duration prior to this latest date: one using pages published within the latest 1 day as new pages, and the other using pages published within the latest 30 days as new pages.

\begin{table*}[t]
	\centering
	\caption{Data statistics: The number of index pages, content pages, the total number of collected pages (fixed at 10,000 pages), the number of pages published within latest 1 day, and the number of pages published within latest 30 days for each site.}
	\scalebox{1.00}{
		{\tabcolsep=4.0pt
			\begin{tabular}{llccccc} \hline
				Category & Site Name & \#Index & \#Content & \#Total & \#Latest 1 Day & \#Latest 30 Days \\ \hline
				\multirow{2}{*}{Dev} & CNN & 2,811 & 7,189 & 10,000 & 165 & 1,470 \\
				& Variety & 3,924 & 6,076 & 10,000 & 177 & 1,429 \\ \hline
				\multirow{3}{*}{Test} & TechCrunch & 3,721 & 6,279 & 10,000 & 44 & 763 \\
				& Mongabay & 3,911 & 6,089 & 10,000 & 13 & 150 \\
				& Space.com & 1,964 & 8,036 & 10,000 & 19 & 316 \\ \hline
				\multirow{4}{*}{Noisy-Test} & Entertainment Weekly & -- & -- & 10,000 & 74 & 722 \\
				& The New York Times & -- & -- & 10,000 & 419 & 2,073 \\
				& MedicalNewsToday & -- & -- & 10,000 & 11 & 128 \\
				& Healthline & -- & -- & 10,000 & 21 & 247 \\ \hline
			\end{tabular}
		}
	}
	\label{tab:dataset_statistics}
\end{table*}

\subsection{Page Type Classification using LLMs}
\label{sec:llm_classification}

A prompt containing the target web page's information and a description of the task is provided to an LLM, which then classifies the input web page as either an index page or a content page. We compare two LLMs offered via the OpenAI API: GPT-4o-mini (gpt-4o-mini-2024-07-18) and GPT-4o (gpt-4o-2024-08-06).

We evaluate two types of input to the LLM: using only the title of the web page, and using both the title and body text of the web page. The body text of web pages is extracted from HTML content using heuristic rules.\footnote{We use a script modified from ExtractContent3 (\url{https://github.com/kanjirz50/python-extractcontent3}).} The title succinctly represents the page's content and serves as a crucial clue for classification. The body text provides more information than the title, and if the LLM can interpret its content sufficiently, it can serve as strong grounds for judgment. However, the extracted body text also carries the risk of including noise, such as titles or summaries from related pages, potentially leading to misclassifications. To determine the optimal input for the LLM, we conduct a comparison of these two approaches.

\begin{table}[t]
	\centering
	\caption{Evaluation results of the page type classification performance on the test set. Reports precision, recall, and F1 score, considering index pages as positive.}
	\scalebox{0.85}{
		{\tabcolsep=4.0pt
			\begin{tabular}{llccc} \hline
				Method & Input & Precision & Recall & F1 \\ \hline
				All Pages & -- & 0.320 & \textbf{1.000} & 0.478 \\
				Rule-based & -- & 0.699 & 0.910 & 0.787 \\
				\multirow{2}{*}{GPT-4o-mini} & Title & 0.989 & 0.643 & 0.777 \\
				& Title + Body & \textbf{0.992} & 0.570 & 0.697 \\
				\multirow{2}{*}{GPT-4o} & Title & 0.980 & 0.734 & 0.836 \\
				& Title + Body & 0.984 & 0.820 & \textbf{0.894} \\ \hline
			\end{tabular}
		}
	}
	\label{tab:result_classification}
\end{table}

\begin{table*}[t]
	\centering
	\caption{Evaluation results of coverage for new pages on the test set. Reports the coverage and their averages for 10 combinations: two settings for new pages (pages published within latest 1 day and pages published within latest 30 days) and five settings for the number of index pages at shallow hierarchy levels used as starting points (top 10, 30, 100, 300, and 1,000 pages).}
	\scalebox{0.80}{
		{\tabcolsep=4.0pt
			\begin{tabular}{llccccccccccc} \hline
				\multirow{2}{*}{Method} & \multirow{2}{*}{Input} & \multicolumn{5}{c}{Latest 1 Day} & \multicolumn{5}{c}{Latest 30 Days} & \multirow{2}{*}{Average} \\
				& & 10P & 30P & 100P & 300P & 1,000P & 10P & 30P & 100P & 300P & 1,000P & \\ \hline
				All Pages & -- & 0.920 & 0.927 & 0.935 & 0.942 & 0.967 & 0.542 & 0.613 & 0.718 & 0.929 & 0.986 & 0.848 \\
				Rule-based & -- & 0.920 & 0.927 & 0.935 & 0.935 & 0.970 & 0.542 & 0.626 & 0.883 & 0.934 & 0.992 & 0.866 \\
				\multirow{2}{*}{GPT-4o-mini} & Title & 0.920 & 0.927 & 0.935 & \textbf{0.970} & 0.970 & 0.542 & \textbf{0.668} & 0.900 & \textbf{0.980} & 0.988 & \textbf{0.880} \\
				& Title + Body & 0.594 & 0.601 & 0.601 & 0.842 & 0.842 & 0.428 & 0.562 & 0.820 & 0.931 & 0.948 & 0.717 \\
				\multirow{2}{*}{GPT-4o} & Title & 0.920 & 0.927 & 0.935 & \textbf{0.970} & 0.970 & 0.542 & 0.643 & \textbf{0.919} & 0.963 & 0.989 & 0.877 \\
				& Title + Body & 0.920 & 0.927 & 0.935 & 0.952 & 0.970 & 0.542 & 0.640 & 0.914 & 0.964 & 0.989 & 0.875 \\
				GPT-4o-mini + All Pages & Title & 0.920 & 0.927 & 0.935 & 0.952 & \textbf{0.977} & 0.542 & 0.633 & 0.880 & 0.952 & \textbf{0.993} & 0.871 \\
				GPT-4o + All Pages & Title + Body & 0.920 & 0.927 & 0.935 & 0.952 & \textbf{0.977} & 0.542 & 0.628 & 0.885 & 0.934 & \textbf{0.993} & 0.869 \\ \hline
				Gold Labels & -- & 0.920 & 0.927 & 0.935 & 0.960 & 0.985 & 0.551 & 0.665 & 0.895 & 0.939 & 0.994 & 0.877 \\ \hline
			\end{tabular}
		}
	}
	\label{tab:result_coverage}
\end{table*}

\begin{table*}[t]
	\centering
	\caption{Evaluation results of coverage for new pages on the noisy-test set. The evaluation method is the same as that described in Table~\ref{tab:result_coverage}.}
	\scalebox{0.80}{
		{\tabcolsep=4.0pt
			\begin{tabular}{llccccccccccc} \hline
				\multirow{2}{*}{Method} & \multirow{2}{*}{Input} & \multicolumn{5}{c}{Latest 1 Day} & \multicolumn{5}{c}{Latest 30 Days} & \multirow{2}{*}{Average} \\
				& & 10P & 30P & 100P & 300P & 1,000P & 10P & 30P & 100P & 300P & 1,000P & \\ \hline
				All Pages & -- & 0.413 & 0.468 & 0.570 & \textbf{0.895} & \textbf{1.000} & 0.393 & 0.445 & 0.593 & \textbf{0.869} & \textbf{0.992} & 0.664 \\
				Rule-based & -- & 0.390 & 0.464 & 0.563 & 0.833 & 0.925 & 0.390 & 0.448 & 0.593 & 0.851 & 0.943 & 0.640 \\
				\multirow{2}{*}{GPT-4o-mini} & Title & 0.413 & 0.468 & 0.700 & 0.792 & 0.968 & \textbf{0.424} & \textbf{0.499} & 0.750 & 0.831 & 0.941 & 0.678 \\
				& Title + Body & 0.216 & 0.317 & 0.626 & 0.656 & 0.963 & 0.362 & 0.445 & 0.704 & 0.722 & 0.953 & 0.596 \\
				\multirow{2}{*}{GPT-4o} & Title & 0.413 & 0.445 & \textbf{0.728} & 0.833 & 0.972 & 0.381 & 0.466 & 0.759 & 0.823 & 0.938 & 0.676 \\
				& Title + Body & 0.413 & \textbf{0.472} & \textbf{0.728} & 0.819 & 0.970 & 0.394 & 0.454 & 0.758 & 0.824 & 0.948 & 0.678 \\
				GPT-4o-mini + All Pages & Title & 0.413 & 0.456 & 0.725 & 0.874 & 0.977 & 0.393 & 0.488 & \textbf{0.760} & 0.848 & 0.956 & \textbf{0.689} \\
				GPT-4o + All Pages & Title + Body & 0.413 & \textbf{0.472} & \textbf{0.728} & 0.866 & 0.977 & 0.393 & 0.453 & \textbf{0.760} & 0.842 & 0.962 & 0.687 \\ \hline
			\end{tabular}
		}
	}
	\label{tab:result_coverage_noisy}
\end{table*}

\section{Experiments}
\label{sec:experiment}

\subsection{Evaluation Methods}
\noindent\textbf{Evaluation of Page Classification Performance}
The predictions of each method are evaluated by directly comparing them with the gold labels obtained through automated annotation. Treating index pages as positive, we calculate the classification precision, recall, and F1 score for each site and report the average values.

\noindent\textbf{Evaluation of New Page Coverage}
We evaluate how comprehensively new pages can be collected by starting from the pages identified as index pages by each method. Coverage is defined as the proportion of new pages that can be reached either directly through links on index pages or indirectly via other new pages, divided by the total number of new pages. For example, if a website has 100 new pages and 80 of them can be reached from the index pages, the coverage is 0.80. The reason why pages that can be reached via new pages are also considered ``reachable'' is that in actual crawling scenarios, the HTML content of newly accessed pages is also analyzed, and if they contain links to new pages, those pages can also be accessed. The coverage is calculated for each website, and the average value is reported.

Additionally, to evaluate coverage under the same resource constraints for each method, we report the coverage when starting from a fixed number of shallow hierarchy index pages determined by each method.\footnote{Shallow hierarchy pages mean pages that can be accessed with fewer transitions from the home page.} We report the coverage for five cases of 10, 30, 100, 300, and 1,000 fixed index pages used. In cases where the number of index pages is fewer than the fixed number of pages used for evaluation, shallow hierarchy content pages are added to maintain consistency in evaluation.

\subsection{Comparison Methods}
We compare the performance of the following methods:

\begin{itemize}[noitemsep, topsep=1.0pt]
	\setlength{\parskip}{1.0pt}
	\setlength{\itemsep}{1.0pt}
	
	\item \textbf{All Pages}: All pages are treated as index pages. For the coverage evaluation, a fixed number of the shallowest hierarchy pages are used as starting points.
	
	\item \textbf{Rule-Based}: Pages with 9 or fewer words in their titles are treated as index pages.
	
	\item \textbf{LLM}: The types of pages are classified using LLMs. We evaluate four combinations using two types of LLMs (GPT-4o-mini and GPT-4o) with two types of inputs (title only and title + body).
\end{itemize}

The following methods are evaluated only for new page coverage:

\begin{itemize}[noitemsep, topsep=1.0pt]
	\setlength{\parskip}{1.0pt}
	\setlength{\itemsep}{1.0pt}
	
	\item \textbf{LLM + All Pages}: We evaluate a hybrid method where, for a fixed number of access starting points, half are selected from the pages determined to be index pages by LLM and half are selected from the shallow hierarchy pages regardless of their type.
	
	\item \textbf{Gold Labels}: Annotations created using content listing pages are used.
\end{itemize}

\subsection{Results}
Table~\ref{tab:result_classification} shows the results of the page type classification evaluation. GPT-4o with Title + Body achieves the highest F1 score, outperforming the baseline methods (All Pages and Rule-Based). While all methods using LLMs exhibit high precision, there are significant differences in their recall scores. In other words, LLM-based methods rarely misclassify content pages as index pages, but they tend to misclassify index pages as content pages. Furthermore, while GPT-4o improves its performance using the web page body, GPT-4o-mini's performance deteriorates when the body is included. This is likely because the body text, automatically extracted from HTML, includes noise that is difficult for GPT-4o-mini to process appropriately, such as titles or summaries of related pages, and GPT-4o-mini may lack the capability to effectively interpret the body text.

Next, Table~\ref{tab:result_coverage} presents the results of the new page coverage evaluation on the test set. The results confirm that LLM-based methods generally perform better than baseline methods. Regarding the overall average, there is no significant difference among LLM-based methods (except for GPT-4o-mini with Title + Body, which performs poorly), and the results are nearly equivalent to those obtained using gold labels. This suggests that once classification performance reaches a certain threshold, its correlation with coverage becomes low, and further improvements in coverage would require improvements in other elements such as the crawl algorithm.

When comparing the number of index pages used as starting points, there is no notable difference in coverage among methods for smaller page numbers. However, when using pages published within the latest 1 day as new pages, differences emerge when using 300 or more index pages, and when using pages published within the latest 30 days as new pages, differences are observed when using 30 or more index pages. This is likely because shallow hierarchy pages, especially the home page and pages directly linked from the home page, are often index pages, and when using a small number of index pages as starting points, all methods tend to select the same set of pages around the home page, leading to similar coverage. On the other hand, as the number of index pages used as starting points increases, the page sets selected by each method become different, and the quality of page selection becomes more impactful, resulting in performance differences.

The evaluation results for the noisy-test set, shown in Table~\ref{tab:result_coverage_noisy}, indicate that LLMs achieve higher coverage than baseline methods, particularly when using 100 index pages. However, for 300 or more pages, the All Pages method achieves higher coverage. This is likely because, on the noisy-test set, it is difficult to comprehensively cover new pages using only index pages. When starting from a large number of pages, sequentially accessing from shallow hierarchy pages regardless of their type can be more effective in collecting new pages. While each method has its strengths and weaknesses, hybrid methods consistently achieve high coverage under the evaluated conditions. Notably, GPT-4o-mini + All Pages with Title achieves the best results in terms of the average coverage on the noisy-test set. Furthermore, to reinforce the validity of the experimental results, the results of additional experiments using a dataset reconstructed at a different time from the same websites are shown in Appendix~\ref{sec:result_reconstructed}.

\section{Conclusion}
\label{sec:conclusion}

In this study, we proposed a method to improve crawling efficiency by classifying web pages into two types, index pages and content pages, using LLMs, and then using the classification results to crawl web pages by starting from index pages. In the experiments, we constructed a dataset with labeled page types and evaluated the method from two perspectives: page type classification performance and coverage for new pages. The results confirmed that LLMs achieve high performance in both evaluation criteria. Future challenges include the following:

\begin{itemize}[noitemsep,topsep=1.0pt] 
	\setlength{\parskip}{1.0pt}
	\setlength{\itemsep}{1.0pt}
	
	\item \textbf{Subdividing Web Page Types}: In this study, pages were classified simply into index pages and content pages, but it may be possible to improve the efficiency of page crawling by further subdividing them. For example, index pages could be divided into pages that contain links to new pages and pages that contain links to old pages, allowing access to be concentrated on the former to retrieve new pages.
	
	\item \textbf{Revisiting Content Pages}: This study focused on collecting new pages, but it is also important to revisit important content pages during crawling to maintain the freshness of collected web pages. Existing studies have proposed importance measurement by PageRank~\cite{Page1998PageRank} or estimated future update frequencies based on each page's update frequency~\cite{10.1145/775152.775246}, but these approaches encounter a cold-start problem where it is difficult for them to handle new pages. A potential approach is to initially estimate the importance or future update frequencies of new pages using LLMs, and then adjust these estimates using PageRank or actual update frequencies after a certain period of time.	
	
	\item \textbf{Validation of Computational Cost}: Since page classification with LLMs incurs a certain computational cost, it is necessary to verify whether the improvement in collection efficiency justifies the increase in computational costs.
	
	\item \textbf{Lightweight LLMs}: Further improvements in classification performance may not directly enhance coverage performance. Therefore, achieving sufficient classification performance with lightweight LLMs will be a key challenge.
	
	\item \textbf{Practical evaluation}: The evaluation in this study is limited to a small number of news websites, so it is necessary to conduct evaluations targeting a wider variety of websites. In addition, rather than evaluations using static snapshots of websites as in this study, evaluations in actual environments where website structures gradually change are also necessary.
	
\end{itemize}

\bibliography{reference}

\appendix

\section{Experiments with the reconstructed dataset}
\label{sec:result_reconstructed}

The experimental results reported in this study were obtained using web pages collected around January 2025. To reinforce the validity of our experimental results, we recollected web pages from the same websites using breadth-first search starting from the home page around March 2025 and conducted the same experiments using the same procedures. Table~\ref{tab:dataset_statistics_reconstructed} shows the statistics of the reconstructed dataset, and Tables~\ref{tab:result_classification_reconstructed}, ~\ref{tab:result_coverage_reconstructed}, and ~\ref{tab:result_coverage_noisy_reconstructed} show the experimental results. These results exhibit the same trends as the results reported in Section~\ref{sec:experiment}, indicating that our method is robust against changes in site structure and web page content over a certain period of time.

\begin{table*}[t]
	\centering
	\caption{Data statistics of the reconstructed dataset.}
	\scalebox{1.0}{
		{\tabcolsep=4.0pt
			\begin{tabular}{llccccc} \hline
				Category & Site Name & \#Index & \#Content & \#Total & \#Latest 1 Day & \#Latest 30 Days \\ \hline
				\multirow{2}{*}{Dev} & CNN & 2,216 & 7,784 & 10,000 & 171 & 1,709 \\
				& Variety & 3,925 & 6,075 & 10,000 & 153 & 1,662 \\ \hline
				\multirow{3}{*}{Test} & TechCrunch & 3,230 & 6,770 & 10,000 & 69 & 911 \\
				& Mongabay & 3,918 & 6,082 & 10,000 & 16 & 217 \\
				& Space.com & 2,549 & 7,451 & 10,000 & 19 & 307 \\ \hline
				\multirow{4}{*}{Noisy-Test} & Entertainment Weekly & -- & -- & 10,000 & 75 & 942 \\
				& The New York Times & -- & -- & 10,000 & 438 & 2,301 \\
				& MedicalNewsToday & -- & -- & 10,000 & 14 & 144 \\
				& Healthline & -- & -- & 10,000 & 19 & 307 \\ \hline
			\end{tabular}
		}
	}
	\label{tab:dataset_statistics_reconstructed}
\end{table*}

\begin{table*}[t]
	\centering
	\caption{Evaluation results of the page type classification performance on the reconstructed test set.}
	\scalebox{0.85}{
		{\tabcolsep=4.0pt
			\begin{tabular}{llccc} \hline
				Method & Input & Precision & Recall & F1 \\ \hline
				All Pages & -- & 0.323 & \textbf{1.000} & 0.486 \\
				Rule-based & -- & 0.680 & 0.834 & 0.749 \\
				\multirow{2}{*}{GPT-4o-mini} & Title & 0.989 & 0.561 & 0.716 \\
				& Title + Body & \textbf{0.991} & 0.513 & 0.649 \\
				\multirow{2}{*}{GPT-4o} & Title & 0.977 & 0.652 & 0.777 \\
				& Title + Body & 0.982 & 0.734 & \textbf{0.837} \\ \hline
			\end{tabular}
		}
	}
	\label{tab:result_classification_reconstructed}
\end{table*}

\begin{table*}[t]
	\centering
	\caption{Evaluation results of coverage for new pages on the reconstructed test set.}
	\scalebox{0.80}{
		{\tabcolsep=4.0pt
			\begin{tabular}{llccccccccccc} \hline
				\multirow{2}{*}{Method} & \multirow{2}{*}{Input} & \multicolumn{5}{c}{Latest 1 Day} & \multicolumn{5}{c}{Latest 30 Days} & \multirow{2}{*}{Average} \\
				& & 10P & 30P & 100P & 300P & 1,000P & 10P & 30P & 100P & 300P & 1,000P & \\ \hline
				All Pages & -- & \textbf{0.835} & \textbf{0.844} & 0.864 & 0.931 & 0.968 & 0.461 & 0.551 & 0.685 & 0.953 & 0.986 & 0.808 \\
				Rule-based & -- & 0.801 & 0.811 & 0.867 & 0.912 & 0.966 & 0.461 & 0.612 & 0.898 & 0.978 & 0.995 & 0.830 \\
				\multirow{2}{*}{GPT-4o-mini} & Title & 0.801 & 0.811 & 0.867 & 0.923 & 0.942 & 0.459 & \textbf{0.630} & 0.927 & \textbf{0.991} & 0.995 & 0.834 \\
				& Title + Body & 0.702 & 0.702 & 0.888 & 0.923 & 0.942 & 0.418 & 0.552 & \textbf{0.941} & 0.974 & 0.994 & 0.803 \\
				\multirow{2}{*}{GPT-4o} & Title & 0.801 & 0.811 & 0.867 & 0.923 & 0.942 & 0.461 & 0.614 & 0.937 & 0.988 & 0.995 & 0.834 \\
				& Title + Body & 0.801 & 0.811 & 0.867 & 0.923 & 0.942 & \textbf{0.511} & 0.624 & 0.938 & 0.988 & 0.995 & 0.840 \\
				GPT-4o-mini + All Pages & Title & 0.801 & \textbf{0.844} & \textbf{0.901} & \textbf{0.966} & \textbf{0.986} & 0.469 & 0.619 & 0.882 & 0.974 & 0.995 & \textbf{0.844} \\
				GPT-4o + All Pages & Title + Body & 0.801 & \textbf{0.844} & 0.859 & 0.949 & \textbf{0.986} & \textbf{0.511} & 0.619 & 0.843 & 0.984 & 0.995 & 0.839 \\ \hline
				Gold Labels & -- & 0.835 & 0.844 & 0.859 & 0.931 & 0.986 & 0.450 & 0.546 & 0.827 & 0.985 & 0.996 & 0.826 \\ \hline
			\end{tabular}
		}
	}
	\label{tab:result_coverage_reconstructed}
\end{table*}

\begin{table*}[t]
	\centering
	\caption{Evaluation results of coverage for new pages on the reconstructed noisy-test set.}
	\scalebox{0.80}{
		{\tabcolsep=4.0pt
			\begin{tabular}{llccccccccccc} \hline
				\multirow{2}{*}{Method} & \multirow{2}{*}{Input} & \multicolumn{5}{c}{Latest 1 Day} & \multicolumn{5}{c}{Latest 30 Days} & \multirow{2}{*}{Average} \\
				& & 10P & 30P & 100P & 300P & 1,000P & 10P & 30P & 100P & 300P & 1,000P & \\ \hline
				All Pages & -- & 0.374 & 0.398 & 0.544 & \textbf{0.879} & \textbf{0.982} & 0.385 & 0.420 & 0.582 & \textbf{0.866} & \textbf{0.984} & 0.642 \\
				Rule-based & -- & 0.350 & 0.391 & 0.523 & 0.790 & 0.912 & \textbf{0.389} & 0.420 & 0.585 & 0.844 & 0.938 & 0.614 \\
				\multirow{2}{*}{GPT-4o-mini} & Title & 0.254 & 0.372 & 0.668 & 0.746 & 0.940 & 0.320 & 0.360 & 0.654 & 0.733 & 0.934 & 0.598 \\
				& Title + Body & 0.253 & 0.340 & 0.671 & 0.699 & 0.954 & 0.379 & 0.452 & 0.738 & 0.768 & 0.939 & 0.619 \\
				\multirow{2}{*}{GPT-4o} & Title & 0.374 & 0.398 & \textbf{0.689} & 0.784 & 0.958 & 0.385 & \textbf{0.470} & 0.742 & 0.816 & 0.936 & 0.655 \\
				& Title + Body & 0.374 & 0.398 & 0.671 & 0.766 & 0.961 & 0.385 & 0.421 & \textbf{0.744} & 0.822 & 0.945 & 0.649 \\
				GPT-4o-mini + All Pages & Title & 0.374 & \textbf{0.406} & 0.687 & 0.829 & 0.964 & 0.385 & 0.427 & 0.741 & 0.841 & 0.954 & \textbf{0.661} \\
				GPT-4o + All Pages & Title + Body & 0.374 & 0.398 & \textbf{0.689} & 0.844 & 0.964 & 0.385 & 0.421 & 0.742 & 0.842 & 0.953 & \textbf{0.661} \\ \hline
			\end{tabular}
		}
	}
	\label{tab:result_coverage_noisy_reconstructed}
\end{table*}

\end{document}